\documentclass[twocolumn]{aastex61}

\newcommand{\simon}{\texttt{SiMon}}

\received{XXX}
\revised{XXX}
\accepted{XXX} 

\shorttitle{P. X. Qian et al.}
\shortauthors{P. X. Qian et al.}

\makeatletter
\def\PYGdefault@reset{\let\PYGdefault@it=\relax \let\PYGdefault@bf=\relax%
    \let\PYGdefault@ul=\relax \let\PYGdefault@tc=\relax%
    \let\PYGdefault@bc=\relax \let\PYGdefault@ff=\relax}
\def\PYGdefault@tok#1{\csname PYGdefault@tok@#1\endcsname}
\def\PYGdefault@toks#1+{\ifx\relax#1\empty\else%
    \PYGdefault@tok{#1}\expandafter\PYGdefault@toks\fi}
\def\PYGdefault@do#1{\PYGdefault@bc{\PYGdefault@tc{\PYGdefault@ul{%
    \PYGdefault@it{\PYGdefault@bf{\PYGdefault@ff{#1}}}}}}}
\def\PYGdefault#1#2{\PYGdefault@reset\PYGdefault@toks#1+\relax+\PYGdefault@do{#2}}

\expandafter\def\csname PYGdefault@tok@gd\endcsname{\def\PYGdefault@tc##1{\textcolor[rgb]{0.63,0.00,0.00}{##1}}}
\expandafter\def\csname PYGdefault@tok@gu\endcsname{\let\PYGdefault@bf=\textbf\def\PYGdefault@tc##1{\textcolor[rgb]{0.50,0.00,0.50}{##1}}}
\expandafter\def\csname PYGdefault@tok@gt\endcsname{\def\PYGdefault@tc##1{\textcolor[rgb]{0.00,0.27,0.87}{##1}}}
\expandafter\def\csname PYGdefault@tok@gs\endcsname{\let\PYGdefault@bf=\textbf}
\expandafter\def\csname PYGdefault@tok@gr\endcsname{\def\PYGdefault@tc##1{\textcolor[rgb]{1.00,0.00,0.00}{##1}}}
\expandafter\def\csname PYGdefault@tok@cm\endcsname{\let\PYGdefault@it=\textit\def\PYGdefault@tc##1{\textcolor[rgb]{0.25,0.50,0.50}{##1}}}
\expandafter\def\csname PYGdefault@tok@vg\endcsname{\def\PYGdefault@tc##1{\textcolor[rgb]{0.10,0.09,0.49}{##1}}}
\expandafter\def\csname PYGdefault@tok@vi\endcsname{\def\PYGdefault@tc##1{\textcolor[rgb]{0.10,0.09,0.49}{##1}}}
\expandafter\def\csname PYGdefault@tok@mh\endcsname{\def\PYGdefault@tc##1{\textcolor[rgb]{0.40,0.40,0.40}{##1}}}
\expandafter\def\csname PYGdefault@tok@cs\endcsname{\let\PYGdefault@it=\textit\def\PYGdefault@tc##1{\textcolor[rgb]{0.25,0.50,0.50}{##1}}}
\expandafter\def\csname PYGdefault@tok@ge\endcsname{\let\PYGdefault@it=\textit}
\expandafter\def\csname PYGdefault@tok@vc\endcsname{\def\PYGdefault@tc##1{\textcolor[rgb]{0.10,0.09,0.49}{##1}}}
\expandafter\def\csname PYGdefault@tok@il\endcsname{\def\PYGdefault@tc##1{\textcolor[rgb]{0.40,0.40,0.40}{##1}}}
\expandafter\def\csname PYGdefault@tok@go\endcsname{\def\PYGdefault@tc##1{\textcolor[rgb]{0.53,0.53,0.53}{##1}}}
\expandafter\def\csname PYGdefault@tok@cp\endcsname{\def\PYGdefault@tc##1{\textcolor[rgb]{0.74,0.48,0.00}{##1}}}
\expandafter\def\csname PYGdefault@tok@gi\endcsname{\def\PYGdefault@tc##1{\textcolor[rgb]{0.00,0.63,0.00}{##1}}}
\expandafter\def\csname PYGdefault@tok@gh\endcsname{\let\PYGdefault@bf=\textbf\def\PYGdefault@tc##1{\textcolor[rgb]{0.00,0.00,0.50}{##1}}}
\expandafter\def\csname PYGdefault@tok@ni\endcsname{\let\PYGdefault@bf=\textbf\def\PYGdefault@tc##1{\textcolor[rgb]{0.60,0.60,0.60}{##1}}}
\expandafter\def\csname PYGdefault@tok@nl\endcsname{\def\PYGdefault@tc##1{\textcolor[rgb]{0.63,0.63,0.00}{##1}}}
\expandafter\def\csname PYGdefault@tok@nn\endcsname{\let\PYGdefault@bf=\textbf\def\PYGdefault@tc##1{\textcolor[rgb]{0.00,0.00,1.00}{##1}}}
\expandafter\def\csname PYGdefault@tok@no\endcsname{\def\PYGdefault@tc##1{\textcolor[rgb]{0.53,0.00,0.00}{##1}}}
\expandafter\def\csname PYGdefault@tok@na\endcsname{\def\PYGdefault@tc##1{\textcolor[rgb]{0.49,0.56,0.16}{##1}}}
\expandafter\def\csname PYGdefault@tok@nb\endcsname{\def\PYGdefault@tc##1{\textcolor[rgb]{0.00,0.50,0.00}{##1}}}
\expandafter\def\csname PYGdefault@tok@nc\endcsname{\let\PYGdefault@bf=\textbf\def\PYGdefault@tc##1{\textcolor[rgb]{0.00,0.00,1.00}{##1}}}
\expandafter\def\csname PYGdefault@tok@nd\endcsname{\def\PYGdefault@tc##1{\textcolor[rgb]{0.67,0.13,1.00}{##1}}}
\expandafter\def\csname PYGdefault@tok@ne\endcsname{\let\PYGdefault@bf=\textbf\def\PYGdefault@tc##1{\textcolor[rgb]{0.82,0.25,0.23}{##1}}}
\expandafter\def\csname PYGdefault@tok@nf\endcsname{\def\PYGdefault@tc##1{\textcolor[rgb]{0.00,0.00,1.00}{##1}}}
\expandafter\def\csname PYGdefault@tok@si\endcsname{\let\PYGdefault@bf=\textbf\def\PYGdefault@tc##1{\textcolor[rgb]{0.73,0.40,0.53}{##1}}}
\expandafter\def\csname PYGdefault@tok@s2\endcsname{\def\PYGdefault@tc##1{\textcolor[rgb]{0.73,0.13,0.13}{##1}}}
\expandafter\def\csname PYGdefault@tok@nt\endcsname{\let\PYGdefault@bf=\textbf\def\PYGdefault@tc##1{\textcolor[rgb]{0.00,0.50,0.00}{##1}}}
\expandafter\def\csname PYGdefault@tok@nv\endcsname{\def\PYGdefault@tc##1{\textcolor[rgb]{0.10,0.09,0.49}{##1}}}
\expandafter\def\csname PYGdefault@tok@s1\endcsname{\def\PYGdefault@tc##1{\textcolor[rgb]{0.73,0.13,0.13}{##1}}}
\expandafter\def\csname PYGdefault@tok@ch\endcsname{\let\PYGdefault@it=\textit\def\PYGdefault@tc##1{\textcolor[rgb]{0.25,0.50,0.50}{##1}}}
\expandafter\def\csname PYGdefault@tok@m\endcsname{\def\PYGdefault@tc##1{\textcolor[rgb]{0.40,0.40,0.40}{##1}}}
\expandafter\def\csname PYGdefault@tok@gp\endcsname{\let\PYGdefault@bf=\textbf\def\PYGdefault@tc##1{\textcolor[rgb]{0.00,0.00,0.50}{##1}}}
\expandafter\def\csname PYGdefault@tok@sh\endcsname{\def\PYGdefault@tc##1{\textcolor[rgb]{0.73,0.13,0.13}{##1}}}
\expandafter\def\csname PYGdefault@tok@ow\endcsname{\let\PYGdefault@bf=\textbf\def\PYGdefault@tc##1{\textcolor[rgb]{0.67,0.13,1.00}{##1}}}
\expandafter\def\csname PYGdefault@tok@sx\endcsname{\def\PYGdefault@tc##1{\textcolor[rgb]{0.00,0.50,0.00}{##1}}}
\expandafter\def\csname PYGdefault@tok@bp\endcsname{\def\PYGdefault@tc##1{\textcolor[rgb]{0.00,0.50,0.00}{##1}}}
\expandafter\def\csname PYGdefault@tok@c1\endcsname{\let\PYGdefault@it=\textit\def\PYGdefault@tc##1{\textcolor[rgb]{0.25,0.50,0.50}{##1}}}
\expandafter\def\csname PYGdefault@tok@o\endcsname{\def\PYGdefault@tc##1{\textcolor[rgb]{0.40,0.40,0.40}{##1}}}
\expandafter\def\csname PYGdefault@tok@kc\endcsname{\let\PYGdefault@bf=\textbf\def\PYGdefault@tc##1{\textcolor[rgb]{0.00,0.50,0.00}{##1}}}
\expandafter\def\csname PYGdefault@tok@c\endcsname{\let\PYGdefault@it=\textit\def\PYGdefault@tc##1{\textcolor[rgb]{0.25,0.50,0.50}{##1}}}
\expandafter\def\csname PYGdefault@tok@mf\endcsname{\def\PYGdefault@tc##1{\textcolor[rgb]{0.40,0.40,0.40}{##1}}}
\expandafter\def\csname PYGdefault@tok@err\endcsname{\def\PYGdefault@bc##1{\setlength{\fboxsep}{0pt}\fcolorbox[rgb]{1.00,0.00,0.00}{1,1,1}{\strut ##1}}}
\expandafter\def\csname PYGdefault@tok@mb\endcsname{\def\PYGdefault@tc##1{\textcolor[rgb]{0.40,0.40,0.40}{##1}}}
\expandafter\def\csname PYGdefault@tok@ss\endcsname{\def\PYGdefault@tc##1{\textcolor[rgb]{0.10,0.09,0.49}{##1}}}
\expandafter\def\csname PYGdefault@tok@sr\endcsname{\def\PYGdefault@tc##1{\textcolor[rgb]{0.73,0.40,0.53}{##1}}}
\expandafter\def\csname PYGdefault@tok@mo\endcsname{\def\PYGdefault@tc##1{\textcolor[rgb]{0.40,0.40,0.40}{##1}}}
\expandafter\def\csname PYGdefault@tok@kd\endcsname{\let\PYGdefault@bf=\textbf\def\PYGdefault@tc##1{\textcolor[rgb]{0.00,0.50,0.00}{##1}}}
\expandafter\def\csname PYGdefault@tok@mi\endcsname{\def\PYGdefault@tc##1{\textcolor[rgb]{0.40,0.40,0.40}{##1}}}
\expandafter\def\csname PYGdefault@tok@kn\endcsname{\let\PYGdefault@bf=\textbf\def\PYGdefault@tc##1{\textcolor[rgb]{0.00,0.50,0.00}{##1}}}
\expandafter\def\csname PYGdefault@tok@cpf\endcsname{\let\PYGdefault@it=\textit\def\PYGdefault@tc##1{\textcolor[rgb]{0.25,0.50,0.50}{##1}}}
\expandafter\def\csname PYGdefault@tok@kr\endcsname{\let\PYGdefault@bf=\textbf\def\PYGdefault@tc##1{\textcolor[rgb]{0.00,0.50,0.00}{##1}}}
\expandafter\def\csname PYGdefault@tok@s\endcsname{\def\PYGdefault@tc##1{\textcolor[rgb]{0.73,0.13,0.13}{##1}}}
\expandafter\def\csname PYGdefault@tok@kp\endcsname{\def\PYGdefault@tc##1{\textcolor[rgb]{0.00,0.50,0.00}{##1}}}
\expandafter\def\csname PYGdefault@tok@w\endcsname{\def\PYGdefault@tc##1{\textcolor[rgb]{0.73,0.73,0.73}{##1}}}
\expandafter\def\csname PYGdefault@tok@kt\endcsname{\def\PYGdefault@tc##1{\textcolor[rgb]{0.69,0.00,0.25}{##1}}}
\expandafter\def\csname PYGdefault@tok@sc\endcsname{\def\PYGdefault@tc##1{\textcolor[rgb]{0.73,0.13,0.13}{##1}}}
\expandafter\def\csname PYGdefault@tok@sb\endcsname{\def\PYGdefault@tc##1{\textcolor[rgb]{0.73,0.13,0.13}{##1}}}
\expandafter\def\csname PYGdefault@tok@k\endcsname{\let\PYGdefault@bf=\textbf\def\PYGdefault@tc##1{\textcolor[rgb]{0.00,0.50,0.00}{##1}}}
\expandafter\def\csname PYGdefault@tok@se\endcsname{\let\PYGdefault@bf=\textbf\def\PYGdefault@tc##1{\textcolor[rgb]{0.73,0.40,0.13}{##1}}}
\expandafter\def\csname PYGdefault@tok@sd\endcsname{\let\PYGdefault@it=\textit\def\PYGdefault@tc##1{\textcolor[rgb]{0.73,0.13,0.13}{##1}}}

\makeatother

\usepackage{listings}
\usepackage[draft]{minted}
\usepackage{CJKutf8}
\begin{document}

\title[Simulation Monitor for Computational Astrophysics]{SiMon: Simulation Monitor for Computational Astrophysics}

\correspondingauthor{Penny Xuran Qian}
\email{quanxuran@nao.cas.cn}

\author{Penny Xuran Qian (\begin{CJK*}{UTF8}{gbsn}钱旭冉\end{CJK*})}
\affiliation{National Astronomical Observatories, Chinese Academy of Sciences, Beijing 100012, China}
\affiliation{Harvard-Smithsonian Center for Astrophysics, Cambridge 02138, MA, U.S.A.}

\author{Maxwell Xu Cai (\begin{CJK*}{UTF8}{gbsn}蔡栩\end{CJK*})}
\affiliation{Leiden Observatory, Leiden University, PO Box 9513, 2300 RA, Leiden, The Netherlands}

\author{Simon Portegies Zwart}
\affiliation{Leiden Observatory, Leiden University, PO Box 9513, 2300 RA, Leiden, The Netherlands}

\author{Ming Zhu (\begin{CJK*}{UTF8}{gbsn}朱明\end{CJK*})}
\affiliation{National Astronomical Observatories, Chinese Academy of Sciences, Beijing 100012, China}


\begin{abstract}
Scientific discovery via numerical simulations is important in modern astrophysics. This relatively new branch of astrophysics has become possible due to the development of reliable numerical algorithms and the high performance of modern computing technologies. These enable the analysis of large collections of observational data and the acquisition of new data via simulations at unprecedented accuracy and resolution. Ideally, simulations run until they reach some pre-determined  termination condition, but often other factors cause extensive numerical approaches to break down at an earlier stage. In those cases,  processes tend to be interrupted due to unexpected events in the  software or the hardware. In those cases, the scientist handles the interrupt manually, which is time-consuming and prone to errors. We present the Simulation Monitor (\simon{}) to automatize the farming of large and extensive simulation processes. Our method is light-weight, it fully automates the entire workflow management, operates concurrently across multiple platforms and can be installed in user space. Inspired by the process of crop farming, we perceive each simulation as a crop in the field and running simulation becomes analogous to growing crops. With the development of \simon{} we relax the technical aspects of simulation management. The initial package was developed  for extensive parameter searchers in numerical simulations, but it turns out to work equally well for
automating the computational processing and reduction of observational data reduction. 
\end{abstract}



\section{Introduction}
Numerical simulations are widely used to investigate dynamical systems on a wide variety of scales. In astrophysics these scales range from planetary systems via open clusters and globular clusters, to galaxies and even large cosmological scales \citep{Trenti:2008aa}. In each scale of the system, multiple astrophysical processes are involved, such as $N$-body system dynamics, radiative transfer, magnetohydrodynamics, stellar evolution and fluid dynamics. Started from the pioneering work of \citealt{1960ZA.....50..184V,1964ApNr....9..313A,1968BAN....19..479V} in the early 1960s using up to 100 particles, computational astrophysics have undergone more than six decades of active development, currently with the capability to handle realistic astrophysical systems with unprecedented scales \citep{Bedorf:2014:PGT:2683593.2683600} and accuracy \citep{2041-8205-785-1-L3}. At present, various integration algorithms have been developed to tackle different astrophysical systems, such as symplectic integrators for long-term evolution of planetary systems \citep[e.g.,][]{Wisdom:1991aa}, Hermite scheme for star cluster dynamics \citep{Makino:1991aa}, adaptive mesh refinement \citep{Berger:1989aa} for cosmological simulation, 
Pair-wise symplectic Kepler based integrators \cite{2012NewA...17..711P} and smoothed particle hydrodynamics solvers for fluid dynamics \citep{Gingold:1977aa}.

While numerical simulations have provided effective approaches to investigate the evolution of chaotic and nonlinear astrophysical systems, the resulting computational efforts are usually extensive. Addressing an astrophysical problem with numerical simulations usually involves a large parameter space, which in turns requires a large number of simulations to be carried out. For example, to compute the cross sections for planetary systems interacting with passing stars and binaries, \citealt{Li:2015aa} carry out over two million individual scattering simulations. As another example, to understand how varying galactic tide affects the dynamical evolution of star clusters, \citealt{Cai:2016aa} perform a grid of simulations for star clusters by exploring the parameter space of a different number of stars and orbital eccentricities. \citealt{Jilkova:2016aa} study the mass transfer between debris discs during close stellar encounters using a grid of 1000 runs. In the cosmological context, it is common to deal with an ultra large number of particles ($N > 10^{11}$)  \citep[e.g.,][]{2010IEEEC..43...63P,Springel:2001aa}, which takes a long time to carry out even on supercomputers.

Ideally, simulations run without interruptions from the moment they are submitted to the computer queue until they reach the termination criteria. In reality, however, simulations tend to be interrupted for a variety of reasons, including regular computer maintenance, job scheduling limitations, power outage or unscheduled interrupts due to problems with the software and hardware. Software problems in a simulation code can be simply bugs, or physics-related such as excessive energy errors or numerical singularities (e.g., tight binary systems, close encounters if two-body relaxation). Hardware problems, such as power failure and disk error can potentially corrupt the output files, making the simulation results unreadable. Also, limitations in the wallclock time of computing clusters imposed by job scheduling systems can terminate simulations prematurely.  The hazards of losing data due to potential software or hardware problems can in principle be circumvented if the simulation snapshots or restart files are backed up sufficiently frequently. In the event of interruptions, human supervision is then required to correct the error, for example by adjusting input parameters and subsequently resubmitting the simulation. Restarting a production simulation is prone to human error. With the ever growing scales and accuracy demands in the research, manual bookkeeping of multiple lengthy simulations becomes increasingly difficult.

While manual bookkeeping is challenging, implementing automatic bookkeeping is rather straightforward. From a technical point of view, a simulation is essentially a process in the underlying operating system. Modern operating systems provide facilities to monitor and control processes. For example, the \texttt{top} command in the UNIX/Linux systems presents an overview of all running processes, memory and CPU usages. In the Information Technology (IT) field, automatic bookkeeping has been sophisticatedly developed into the concept of Application Performance Management (APM), mainly targeting the performance monitoring of end user experience and business transactions. Such tools are usually commercially driven, and their adaptation to monitoring scientific calculations are difficult. In high-performance computing, several open-source job schedulers already exist, such as \texttt{SLURM}\footnote{\href{https://slurm.schedmd.com}{\tt https://slurm.schedmd.com}} \citep{Yoo2003} and \texttt{OpenLava}\footnote{\href{http://www.openlava.org}{\tt http://www.openlava.org}}. Some of these tools also support monitoring server status, but this is different from the need of monitoring the execution of astrophysical calculations. In astronomy, a few tools with the features of monitoring and scheduling have been used to automate the workflow of telescope observations. For example, the SKA\footnote{\href{https://skatelescope.org}{\tt https://skatelescope.org}} Telescope Manager aims at scheduling observations, controlling their execution, monitoring the telescope health status, diagnosing and fixing its faults and so on \citep{2016SPIE.9913E..3SD}, and the JCMT\footnote{\href{https://www.eaobservatory.org/jcmt}{\tt https://www.eaobservatory.org/jcmt}} Observations Management Project \citep{2002ASPC..281..488E, 2004ASPC..314..728D} provides an integrated software architecture and applications to automate all routine tasks associated with flexible scheduling. However, literature search gave very few results about relevant tools available for computational astrophysics. The calculations involved in computational astrophysics require dedicated parameter setting, monitoring and scheduling, and solutions discussed above cannot meet our requirements.

As driven by this motivation, we develop \simon{}, the Simulation Monitor for computational astrophysics, as a response to the rapidly growing demands of automatic astrophysical simulation management. We use a daemon process to periodically check the running processes of simulation codes, and extracting the information from the output files. Furthermore, the daemon process can be used to restart simulations when an interruption is detected automatically, and to backup simulation data at runtime. The primary purpose and strength of this tool are to apply an automated workflow to facilitate the process of carrying out simulations, from generating initial conditions, to monitoring and controlling simulations, until all simulations are completed and the resulting data are properly processed. As such, astronomers only need to specify the initial parameter space and have the workflow to take care of the rest. This is particularly useful for numerical investigations involving large-scale parameter space and/or prolonged simulations or the processing the large numbers of datasets from observations.

In this paper, we introduce the concept of simulation farming as an analogy to simulation monitoring and scheduling in Section~\ref{sec:sim_farming}. Based on the concept of simulation farming, the implementation is detailed in Section~\ref{sec:imple}. A few example applications are presented in Section~\ref{sec:examples}. We finally discuss and summarized in Section~\ref{sec:conclusion}.

\section{Simulation Farming}
\label{sec:sim_farming}
We consider carrying out numerical simulations on a parallel computer analog to farming crops in the field. Crops grow simultaneously in the field to maximize the output; simulations are running in parallel on the computer to minimize the waiting time. The analogy also extends to the life cycle of running an ensemble of simulations:

\textbf{Prepare the soil:} The soil needs to be prepared before it is suitable to grow crops. Likewise, the underlying computing environments need to be configured before carrying out simulations. The preparation includes probing the hardware environment (e.g., determining the available computing, memory and storage resources) and configuring the software environment (e.g., compiling the numerical codes and library dependencies, deploying the job submission scripts on a computing cluster). 

\textbf{Sow the seeds:} Each crop requires a seed to grow. Likewise, each simulation needs a set of initial condition to start. Therefore, in the context of simulation farming, sowing the seeds is to generate the corresponding initial conditions in the parameter space. In order to maintain a clear data structure, the initial conditions and simulation output data of each simulation are contained in a separate directory. Therefore, sowing the seeds also includes the creation of proper directory structure for each simulation.

\textbf{Cultivate the crops:} Cultivation is the act of caring for or growing crops. In the context of simulation farming, growing crops is equivalent to taking scheduling and launching simulations, and taking care of crops is equivalent to monitoring simulations. Moreover, because of the possible interruptions of simulations, the resulting data should be backed up properly, and crashed simulations should be restarted. When a simulation finishes, the freed computational resources should be used to schedule the next simulation.

\textbf{Harvest the results:} When a simulation finishes, the simulation results can be ``harvested'' by processing the data, for example, generating plots and/or converting to the format that the researchers are ready to carry out subsequent data analysis.

\begin{figure}
\centering
\includegraphics[scale=0.11]{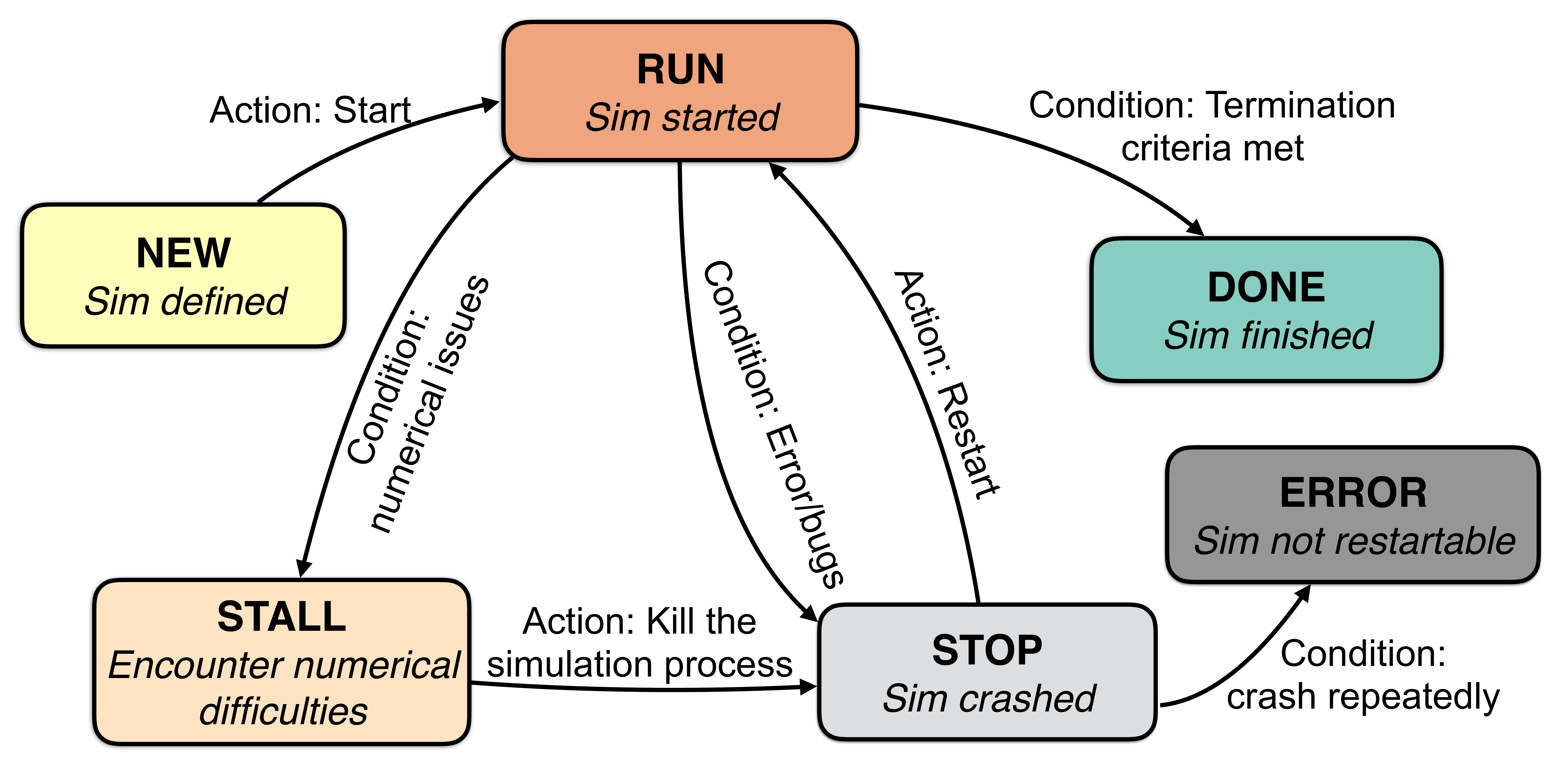}
\caption{The state machine model of the life cycle of a simulation. The life cycle of a simulation can be modeled as a finite state machine, which undergoes the transition of several states. As the code is initialized and initial conditions are loaded, it enters the state of ``NEW''. Subsequently, the code begins to evolve the model and therefore entered the state of ``RUN''. A running simulation may ``STALL'' or ``STOP'' due to various problems. If the simulation undergoes repeated interruptions, it is an indication that there are errors in the code or initial conditions, and human supervision is needed. In such case, the simulation transits from ``STOP'' to ``ERROR'', and human supervision is required. Eventually, the code finishes evolving the mode, and therefore enters the state of ``DONE''. Cycles of new simulations will be triggered unless all simulations are completed.}
\label{fig:sim_state_diagram}
\end{figure}

In addition, just as crops are vulnerable to pests, simulations are vulnerable to software bugs and hardware problems. A simulation may go through the transition from ``NEW'' to ``DONE'', with possible transition to ``RUN'', ``STALL'' and ``STOP''. If a simulation undergoes repeated interruptions, it may transit to the state of ``ERROR'', which requires human supervision. Essentially, the life cycle of each individual simulation can be modeled as a finite state machine, as shown in Fig.~\ref{fig:sim_state_diagram}. Accordingly, the primary objective of simulation farming is to facilitate the transitions of a collection of state machines automatically, each of which as a simulation, from the status of ``NEW'' to the status of ``DONE''. Since minimum human supervision is expected, simulation farming should be handled automatically with a daemon process. In the meanwhile, the users should be able to take manual control when necessary.

\section{Implementation}
\label{sec:imple}
\simon{} is an open-source, \texttt{Python}-based lightweight implementation of automatic simulation farming. The four stages of the farming process (preparing the soil, sowing the seeds, cultivate the crops, and harvesting the grains) are detailed in this section, followed by the discussing of the user interface and extensibility.

\subsection{Preparing the Soil: Configure the Environmental Variables}
Prior to carrying out simulations, \simon{} requires a few global environment variables, configurable through a text configure file \texttt{SiMon.conf}. As the user starts \simon{}, the configure file will be loaded. The following parameters are of particular importance:
\begin{itemize}
\item \texttt{Root\_dir}: the root directory for simulation data storage. \simon{} assumes that each simulation has its own directory for the storage of the initial condition and the resulting data, placed under the root data directory.
\item \texttt{Daemon\_sleep\_time}: the time period at which \simon{} will check through all simulations.
\item \texttt{Max\_concurrent\_jobs}: the number of simulations to be carried out simultaneously.
\item \texttt{Max\_restarts}: the maximum number of times a simulation will be restarted (a simulation is marked as ERROR when exceeding this limit).
\item \texttt{Log\_level}: The level of detail of the logging facility. In decreasing level of severities: \texttt{CRITICAL}, \texttt{ERROR}, \texttt{WARNING}, and \texttt{INFO}. The default log level is \texttt{INFO}, in which all messages are logged.
\end{itemize}

\subsection{Sowing the Seeds: Deploy the Initial Conditions}

To initialize an ensemble of simulations in a given parameter space, we provide facilities to generate the initial conditions and a per-simulation configuration file, and deploy them with a proper structure on the file system. In order to maintain a clear data structure, the initial conditions, configuration files, and the simulation output are contained in separate directories. 
The family of subdirectories are collected in a parent directory, as indicated in the  configuration file. When iterating an ensemble of simulations the workflow parses the per-simulation configuration file to obtain the information regarding how to control the simulation and the priority of the simulation. The  configuration files for each individual simulation are generated automatically by the initial condition generator according to the parameter space specification and global settings. If necessary the user can override the default settings of any simulations by editing the appropriate configuration file.

\subsection{Cultivate the Crops: Automatic Simulation Monitoring and Scheduling}
Monitoring and scheduling are core functionalities, which are automatic and require minimal human supervision. In daemon mode the workflow operates in the background as a service.  An interactive dashboard is provided to control for the simulation in the interactive mode.

\begin{figure*}
	\centering
	\includegraphics[width=\textwidth]{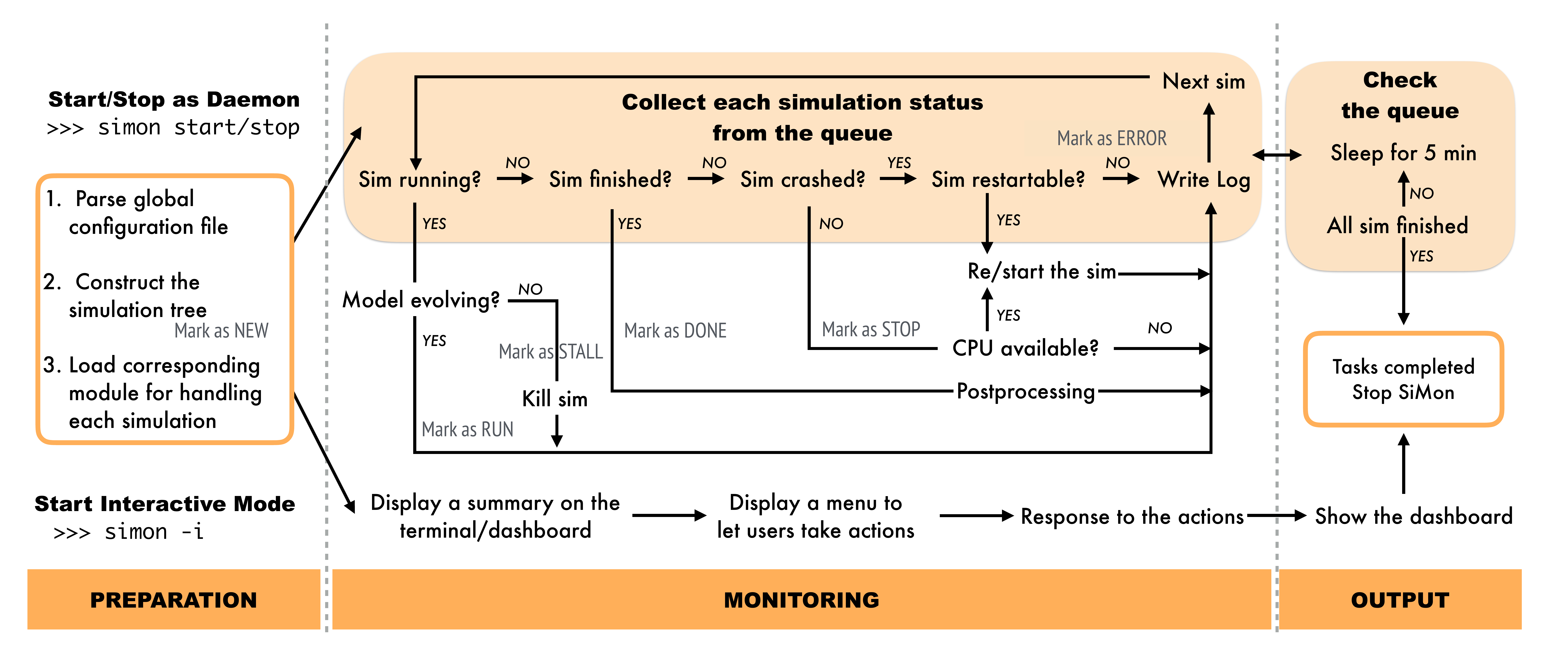}	
	\caption{Flowchart of the daemon mode and the interactive mode of \simon{}. \simon{} maintains a queue of managed simulations, sorted by their priority. The queue is updated either by the daemon periodically or by the time when the user invokes the interactive mode. \simon{} collects the real-time status of all managed simulation, and display the information in its interactive dashboard, or take management actions automatically in the daemon mode.}
	\label{fig:sim_workflow}
\end{figure*}

In Fig~\ref{fig:sim_workflow} we present  the general workflow, which can be divided into three steps:

{\bf Step 1 -- Preparation:} The working directory for simulation is determined from the configuration file.  A beneath-first search (BFS) \citep[see, e.g.,][]{Leiserson:2010:WPB:1810479.1810534} is performed on the simulation data root directory to construct the hierarchical simulation collection. Each simulation task has its own configuration file, which is parsed to determine which code should be employed and loads the corresponding module.  

{\bf Step 2 -- Monitoring:} The input files for starting the run are prepared, as well as output and diagnostics files. In daemon mode, the real-time status of the managed simulations are determined, and management actions are initiated according to the state machine model, as shown in Figure~\ref{fig:sim_state_diagram}. In interactive mode, an information dashboard is presented allowing the users to control the runs manually. 

{\bf Step 3 -- Output:} In interactive mode, an overview of the status of the simulation is performed, and the user can monitor and control the simulation manually. In the daemon mode, each of the automatic actions taken by its scheduling algorithm is logged in a separate file.

\subsubsection{Daemon Mode}
\label{ssec:daemon_mode}

The daemon mode is the key component for automated simulation farming. The underlying idea is shown in the pseudocode below:

\begin{Verbatim}[commandchars=\\\{\}]
\PYGdefault{n}{queue} \PYGdefault{o}{=} \PYGdefault{n}{BFS}\PYGdefault{p}{(}\PYGdefault{n}{parameter} \PYGdefault{n}{space}\PYGdefault{p}{)}
\PYGdefault{k}{while}\PYGdefault{p}{(}\PYGdefault{n}{queue} \PYGdefault{o+ow}{is} \PYGdefault{o+ow}{not} \PYGdefault{n}{empty}\PYGdefault{p}{):}
    \PYGdefault{k}{for} \PYGdefault{n}{sim} \PYGdefault{o+ow}{in} \PYGdefault{n}{queue}\PYGdefault{p}{:}
        \PYGdefault{k}{if} \PYGdefault{n}{sim} \PYGdefault{o+ow}{is} \PYGdefault{n}{running}\PYGdefault{p}{:}
            \PYGdefault{k}{if} \PYGdefault{o+ow}{not} \PYGdefault{n}{sim} \PYGdefault{o+ow}{is} \PYGdefault{n}{evolving}\PYGdefault{p}{:}
                \PYGdefault{n}{Kill}\PYGdefault{p}{(}\PYGdefault{n}{sim}\PYGdefault{p}{)}
                \PYGdefault{n}{Mark}\PYGdefault{p}{(}\PYGdefault{n}{sim}\PYGdefault{p}{,} \PYGdefault{n}{STALL}\PYGdefault{p}{)}
        \PYGdefault{k}{else} \PYGdefault{k}{if} \PYGdefault{n}{sim} \PYGdefault{o+ow}{is} \PYGdefault{n}{finished}\PYGdefault{p}{:}
            \PYGdefault{n}{Mark}\PYGdefault{p}{(}\PYGdefault{n}{sim}\PYGdefault{p}{,} \PYGdefault{n}{DONE}\PYGdefault{p}{)}
            \PYGdefault{n}{Dequeue}\PYGdefault{p}{(}\PYGdefault{n}{sim}\PYGdefault{p}{,} \PYGdefault{n}{queue}\PYGdefault{p}{)}
            \PYGdefault{n}{Finalize}\PYGdefault{p}{(}\PYGdefault{n}{sim}\PYGdefault{p}{)}
        \PYGdefault{k}{else} \PYGdefault{k}{if} \PYGdefault{n}{sim} \PYGdefault{o+ow}{is} \PYGdefault{n}{crashed}\PYGdefault{p}{:}
            \PYGdefault{k}{if} \PYGdefault{n}{sim} \PYGdefault{o+ow}{is} \PYGdefault{n}{restartable}\PYGdefault{p}{:}
                \PYGdefault{k}{if} \PYGdefault{n}{CPU} \PYGdefault{o+ow}{is} \PYGdefault{n}{available}\PYGdefault{p}{:}
                    \PYGdefault{n}{Restart}\PYGdefault{p}{(}\PYGdefault{n}{sim}\PYGdefault{p}{)}
                \PYGdefault{k}{else}\PYGdefault{p}{:}
                    \PYGdefault{n}{Mark}\PYGdefault{p}{(}\PYGdefault{n}{sim}\PYGdefault{p}{,} \PYGdefault{n}{STOP}\PYGdefault{p}{)}
            \PYGdefault{k}{else}\PYGdefault{p}{:}
                \PYGdefault{n}{Mark}\PYGdefault{p}{(}\PYGdefault{n}{sim}\PYGdefault{p}{,} \PYGdefault{n}{ERROR}\PYGdefault{p}{)}
                \PYGdefault{n}{Dequeue}\PYGdefault{p}{(}\PYGdefault{n}{sim}\PYGdefault{p}{,} \PYGdefault{n}{queue}\PYGdefault{p}{)}
                \PYGdefault{n}{GenerateWarning}\PYGdefault{p}{(}\PYGdefault{n}{sim}\PYGdefault{p}{)}
        \PYGdefault{k}{else} \PYGdefault{k}{if} \PYGdefault{n}{sim} \PYGdefault{o+ow}{not} \PYGdefault{n}{crashed}\PYGdefault{p}{:}
            \PYGdefault{k}{if} \PYGdefault{n}{CPU} \PYGdefault{o+ow}{is} \PYGdefault{n}{available}\PYGdefault{p}{:}
                \PYGdefault{n}{Start}\PYGdefault{p}{(}\PYGdefault{n}{sim}\PYGdefault{p}{)}
            \PYGdefault{k}{else}\PYGdefault{p}{:}
                \PYGdefault{n}{Mark}\PYGdefault{p}{(}\PYGdefault{n}{sim}\PYGdefault{p}{,} \PYGdefault{n}{NEW}\PYGdefault{p}{)}
        \PYGdefault{n}{Write}\PYGdefault{p}{(}\PYGdefault{n}{Log} \PYGdefault{n}{files}\PYGdefault{p}{)}
    \PYGdefault{n}{Sleep}\PYGdefault{p}{(}\PYGdefault{n}{a} \PYGdefault{n}{period} \PYGdefault{n}{of} \PYGdefault{n}{time}\PYGdefault{p}{)}
\PYGdefault{n}{Quit}\PYGdefault{p}{(}\PYGdefault{n}{SiMon}\PYGdefault{p}{)}
\end{Verbatim}

Here we check each simulation and collect status information. Subsequent actions are taken according to the state machine model. After this, the manager is put to sleep for the duration of that particular action. This procedure is repeated until each simulation has finished.

The daemon mode facilitates the following tasks:\\

\textbf{Automatic Backup of Simulation Data:} Simulation codes have to support restarting, allowing the user to continue a simulation that was previously interrupted. This is implemented by periodically storing a realization of the simulation data; i.e. a restart file.
Two subsequent restart files are always kept to guarantee that even an undesirable interrupt during the writing of a restart file would not prevent the run to be restarted at a later instance.

\begin{figure*}
\centering
\includegraphics[width=\textwidth]{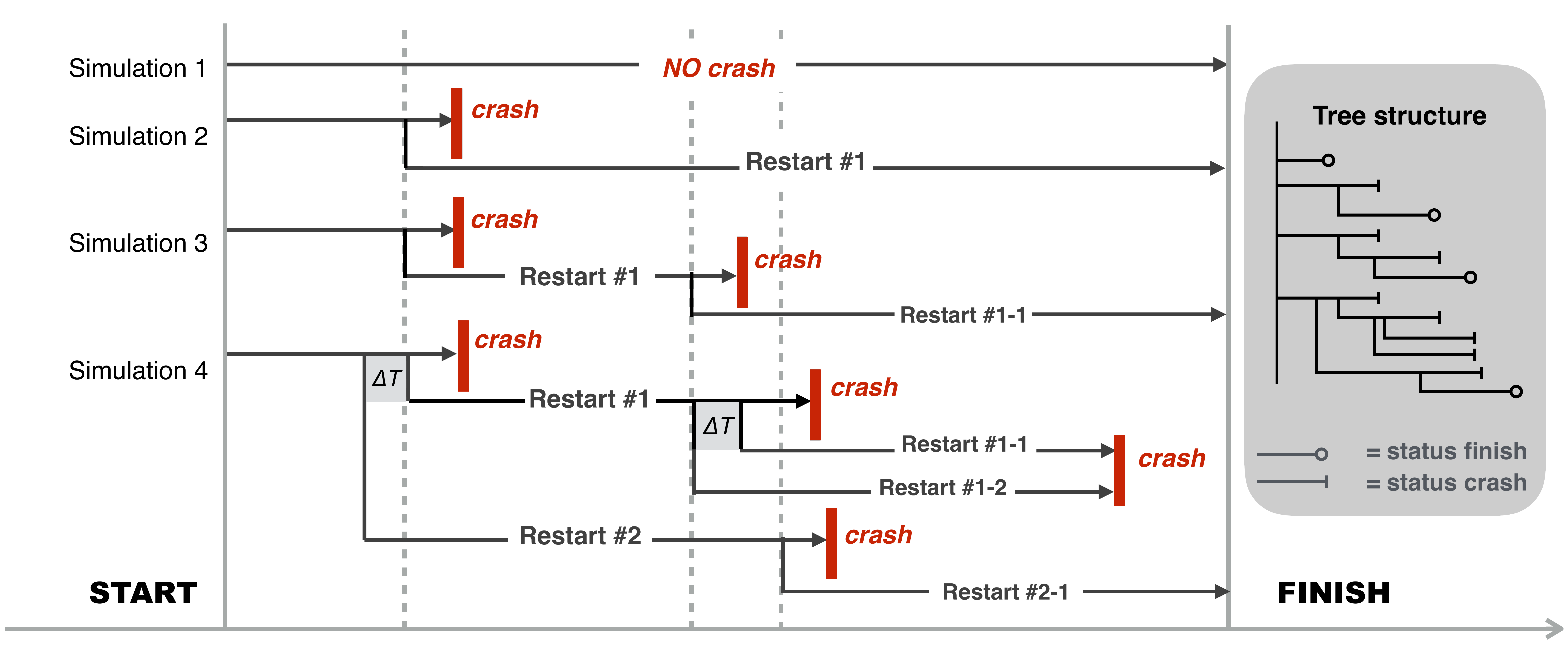}	
\caption{\simon{} uses a hierarchal topology to index an ensemble of simulations. Initially, each simulation is a leaf node under the root node. As the simulations evolve, some of them may crash and therefore be restarted subsequently. The restarted simulation is considered as a child node of the original simulation. Accordingly, the status of the original simulation is obtained through propagating the information of the restarted one. Controlling the original simulation is essentially controlling the restarted simulation in the child node. In this way, the simulation tree grows dynamically until all simulations are finished. 
In this example, we manage an ensemble of four simulations. Simulation number 1 finishes without interruption: no restart is needed. Simulation number 2 stalled at a specific moment but manages to end after one restart \texttt{Restart \#1}. Simulation number 3 crashes (for the second time) after the first restart (\texttt{Restart \#1}), which is named \texttt{Restart \#1-1} in the figure. After these restarts, the code finishes successfully. 
Simulation number 4 has a more complicated evolution. In this case, \emph{rollback restart} is needed to finish the run. After the first interruption, the simulation is restarted as \texttt{Restart \#1}, which subsequently crashed again and restarted as \texttt{Restart \#1-1}. This second restart does not manage to advance in time due to numerical problems, and a subsequent restart is initiated by the scheduler. This subsequent restart adopts a rolling back from the previous snapshot in order to prevent the earlier encountered numerical problems. This new restart is called \texttt{Restart \#1-2} encounters another numerical problem which prevents the simulation to make sufficient progress. By this time all the stored snapshots since \texttt{Restart \#1} are used to restart the simulation, but none of them was successful in progressing the simulation.  The  file \texttt{Restart \#1} is identified as non-restartable. The mitigation action taken by the system is to initiate a  rollback to an earlier time, $\Delta T$ before \texttt{Restart \#1} and restarts from there. This restart file is called \texttt{Restart \#2}. The simulation commences from this restart file after one further interruption, which is resolved by a restart from (\texttt{Restart \#2-1}). The restart tree for this example is presented on the right.}
\label{fig:sim_tree}
\end{figure*}

\textbf{Automatic Restart:} We employ a rollback scheme to restart simulations automatically. When a simulation is interrupted, the most recent restart files will be used. If the last restart files turned out to be problematic for restarting (e.g., corrupted), then the second to last restart file will be used instead (rollback). Each simulation has an ancestor node (the simulation from which it restarts), and several children nodes (simulations from which was restarted). Topologically speaking, the rollback restart scheme forms a tree structure in which the current simulation is a branch.  A schematic example of automatic simulation restarting is shown in Fig. \ref{fig:sim_tree}, where the rollback scheme is illustrated as Simulation 4, and Simulations 1-3 in the figure shows real degenerated trees.

Information about the state of a simulation propagates through the tree nodes. When a simulation is interrupted, the restart tree is searched starting at the initial (root) simulation. When present, the search will commence through all restarts (leaf nodes), until a node with the largest model time is reached. This oldest node is the restart candidate. Upon restart, its state is set to ``RUN'' to indicate that the original simulation is running through one of its restarted simulations. If a simulation undergoes repeated restarts and terminates with the state ``ERROR'', this information is propagated to all its child nodes. This branch in the restart tree will subsequently be terminated.

\begin{figure*}
\centering
\includegraphics[width=0.8\textwidth]{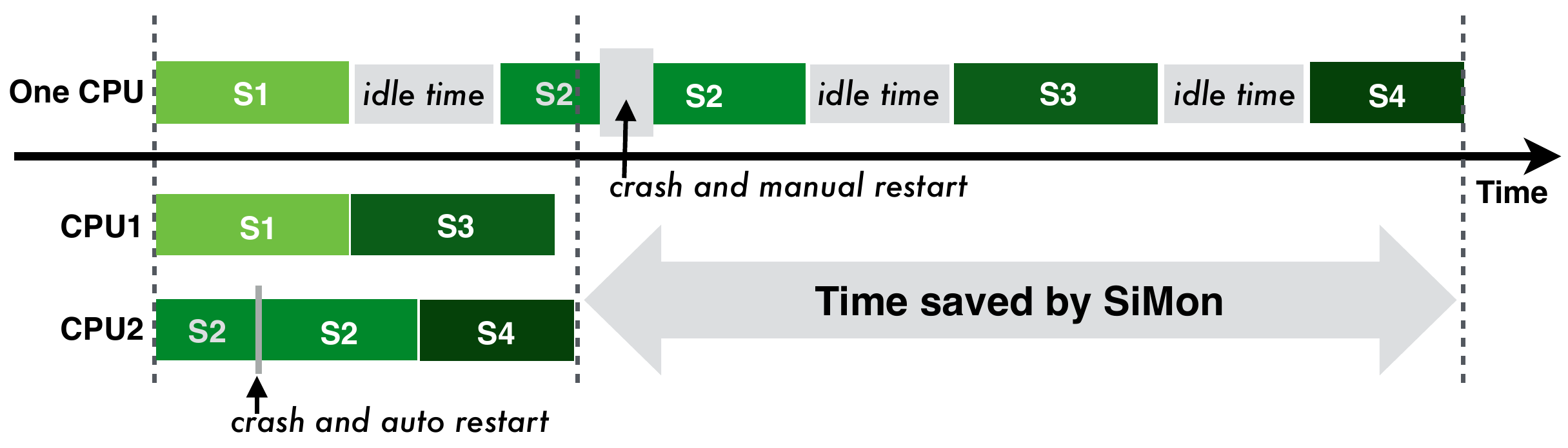}
\caption{The priority based job scheduling scheme of \simon{}, compared with manual sequential simulation management. S1, S2, S3 and S4 are four simulations. The times requires to finish them are indicated with their lengths; the job priorities are indicated with the color (lighter colors ones have higher priorities). The four simulations are launched on a two-CPU machine. In the beginning, both CPU1 and CPU2 are idle, so S1 and S2 are scheduled on them due to their high priorities, respectively. S2 undergoes an interruption, but has been immediately restarted automatically by \simon{}. When S1 is finished, CPU1 becomes idle, and so \simon{} launches S3 immediately. Soon after S3 starts, S2 is finishes, and CPU2 becomes idle, following by the launch of S4. This scheduling scheme significantly reduces the total time of running multiple simulations on multil-processor machines.}
\label{fig:scheduling_scheme}
\end{figure*}

\textbf{Automatic Scheduling:} To maximize the utilization of the available hardware, \simon{} automatically schedules new tasks as soon as computer resources become available. The order in which tasks are scheduled depends on the queue priority, as shown in Figure~\ref{fig:scheduling_scheme}. The daemon regularly checks the status of each simulation, and actions are taken according to the logic detailed in Figure~\ref{fig:sim_state_diagram}. In comparison, if simulations are managed manually, it is difficult for a human user to respond immediately when a simulation is finished or interrupted, and time is wasted when the machine becomes idle. Additionally, it will be even more difficult if the human user attempts to keep all the processors busy at all time (manual parallelization). This priority-based scheduling scheme not only minimizes the idle time, but also adaptively parallelizes the launching of simulations.


\textbf{Automatic Bookkeeping:} Since actions are taken automatically, it is important to record these actions for future reference. Each action performed by the daemon is recorded in the logfile. An example of such a log is
\begin{verbatim}
5/10 0:28AM: sim_1 [INFO] Started.
5/10 1:28AM: sim_2 [INFO] Restarted.
5/10 2:28AM: sim_3 [WARNING] Crashed.
5/10 3:28AM: sim_4 [ERROR] Not restartable.
\end{verbatim}

\subsubsection{Interaction Mode: Manual Control}\label{ssec:inter_mode}
In interactive mode provides the users with an overview of the current status of all simulations, and allow the users to interfere. 

\begin{figure}
\begin{center}
\includegraphics[width=0.45\textwidth]{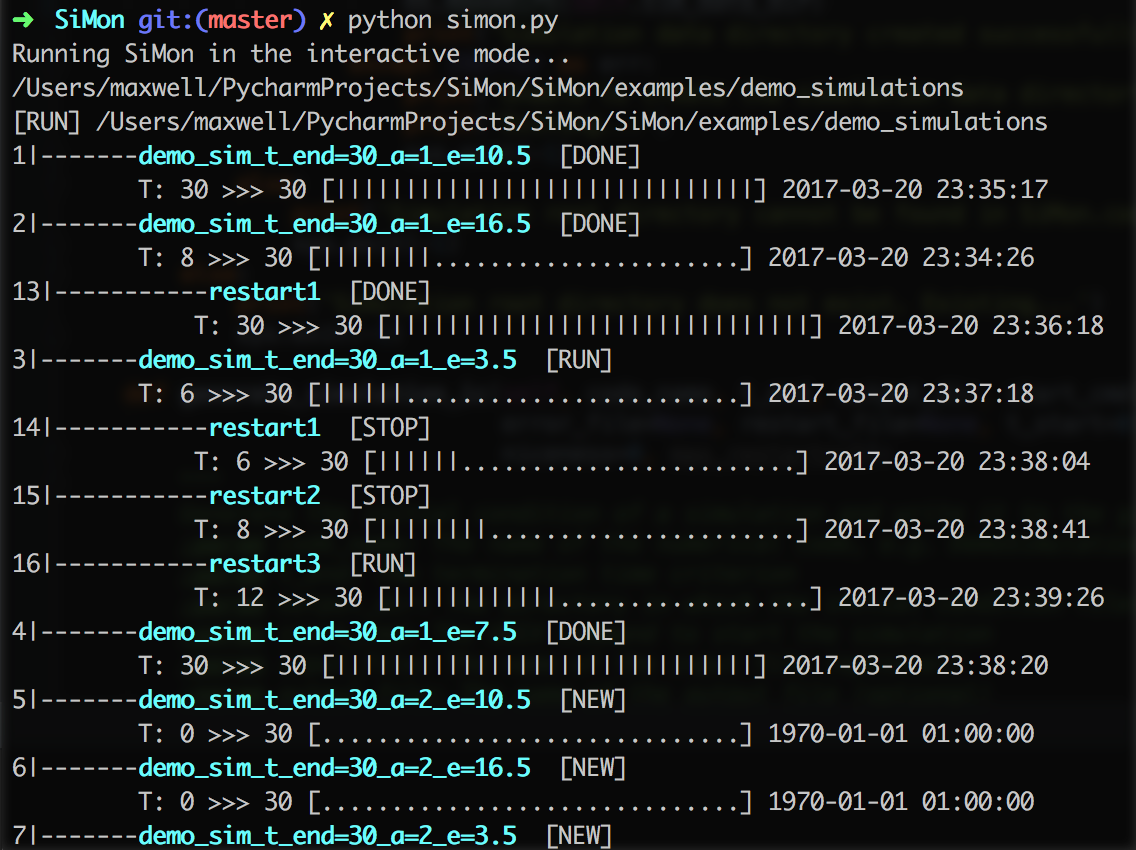}
\caption{The interactive dashboard of \simon{}, which gives the users an overview of the current status of all simulations, and provides facilities for manually control these simulations. Each simulation, including the restarted ones, is assigned with a unique ID, allowing the users to select one or multiple simulations and apply management actions on them. For a list of possible management actions, please refer to Table~\ref{tab:task_list}}.
\label{fig:dashboard}
\end{center}
\end{figure}

The interactive mode is launched from the command-line as {\tt simon}. This interactive mode presents a dashboard with an overview of the real-time status of all simulations (see Fig.\ref{fig:dashboard}). Through this dashboard, the user can select and manipulate the runtime behavior of the simulations (see Fig.~\ref{tab:task_list}). 

\subsection{Harvesting the Grains: Automatic Simulation Data Processing}
When a simulation is finished, the post-processing pipeline is automatically initiated, for example to plot results of informing the user about the status of the run. 

\subsection{User Interface}
The daemon mode of \simon{} has no user interface, as it requires no human supervision. The interactive mode of \simon{} has a text user interface (TUI). This ensures that astronomers can use it without additional rendering support (e.g., graphical user interface, GUI), which is usually the case as they login to a computing node via the Secure Shell (\texttt{SSH}).

\subsection{Extensibility} \label{ssec:extensibility}

\simon{} allows the user to manage an arbitrarily large number of numerical codes, each of which can be distinctively different including the input or post-processing requirements.
In an attempt to generalize the various actions possible, we present a list of the most common actions in 
Table~\ref{tab:sim_attr} and Table.~\ref{tab:sim_func}.  
We implemented a common module that supports any generic simulations environment, this is realized in 
the \texttt{module\_common.py} module. Here we assumed that simulation could be controlled from the \texttt{UNIX} command line or via the configuration file.

\subsection{Availability}
The entire package can be downloaded from GitHub\footnote{\href{https://github.com/maxwelltsai/SiMon}{\tt https://github.com/maxwelltsai/SiMon}}. Alternatively, users can also install using the command \texttt{pip install astrosimon}. The documentation is actively maintained on the package homepage.


\section{Applications}\label{sec:examples}

\subsection{Parameter Space Study: Evolution of Star Clusters on Eccentric Orbits}
We use \simon{} to perform a parameter study of the dynamical evolution of star clusters on eccentric galactic orbits \citep{Cai:2016aa}. In this study, the star clusters are subjected to a periodically varying galactic potential as they orbit the galactic center on different eccentric orbits. They concluded that for those clusters having the same dissolution time the evolution of bound mass and a half-mass radius is approximately independent of their orbital eccentricities. In order to compare different star clusters with the same dissolution time but different galactic orbital eccentricities, the authors iteratively find out the scaling relation between dissolution time and $(a, e, M, m)$ \citep[cf. Sec 2.1 of][]{Cai:2016aa} by varying the mass of the host galaxy ($M$), the mass of the star cluster ($m$), the orbital semi-major axis ($a$) and the eccentricity ($e$) of the cluster's orbit. Each of the simulation in this parameter study takes about a week on a workstation with GPU-acceleration. The simulations themselves are prone to numerical problems due to the dynamical formation of tight binaries, triple/quadruple systems \citep{Aarseth:2003aa}, in which cases the simulation code \texttt{NBODY6} \citep{2012MNRAS.424..545N} may be interrupted by numerical difficulties and therefore needs to be restarted. Without an automated way to address code restarts such a large parameter study would be nearly impossible; we, therefore, performed all calculations for this study using \simon{}.

Prior to the simulations, the method is configured with hardware-specific information (e.g., the numbers of available CPU cores and GPU cores on the computing cluster, the number of CPU cores and GPU cores that can be allocated to a single simulation, the maximum of simulations that can be carried out on the computing cluster concurrently), as well as simulation-specific information (e.g., the Linux commands to start/restart a simulation). Subsequently, we manually deploy initial condition utility to generate a grid of initial conditions (in the term of simulation farming: sow the seeds) and generate a data directory for each. We subsequently start the daemon module which schedules the simulations. Whenever a simulation crashes the output file to probe the problem is parsed and the mitigation strategy deployed. If a simulation stalls, for example, this can happen when tight binaries cause integration steps to approach zero, the code is restarted with a larger time-step parameter. Some simulations are terminated when the relative energy error exceeds a pre-determined limit. In those cases, the code is restarted with a smaller time-step parameter.
Eventually, when all the simulation have finished the data processing pipeline is started automatically, and all results are plotted and stored.

\subsection{Scheduling ensembles of Simulation: Dynamical Evolution of Planetary Systems in Star Clusters}
As a second application, we studied the dynamical evolution of multi-planetary systems in star clusters (Cai et al. 2017, submitted) with numerical simulations, managed by \simon{}. The majority of stars are formed in clustered environments \citep[e.g.,][]{Lada:2003aa}. Planets tend to form within $\sim 10$~Myr following the star formation process \citep[e.g.,][]{Armitage:2007aa}, and because the majority of clusters outlive $\sim 10$\,Myr, planets are typically also formed in star clusters. This was nicely demonstrated also to be the case for the Solar System \citep{2009ApJ...696L..13P}. One can subsequently wonder what the influence of nearby stellar encounters is on the stability and future evolution of planetary systems. The notion that the majority of planetary systems are probably chaotic, small perturbations from an external source could have interesting consequences for the individual planetary systems.
This study is very hard to achieve because of the numerical complexity of these calculations, and the intrinsic chaotic nature of the system. A statistical approach may be one of the most promising routes to understanding these mutual consequences. We carried out such a statistical study using \simon{}, to study the dynamical evolution of star clusters and its planetary systems.

\begin{figure*}
\centering
\includegraphics[scale=0.9]{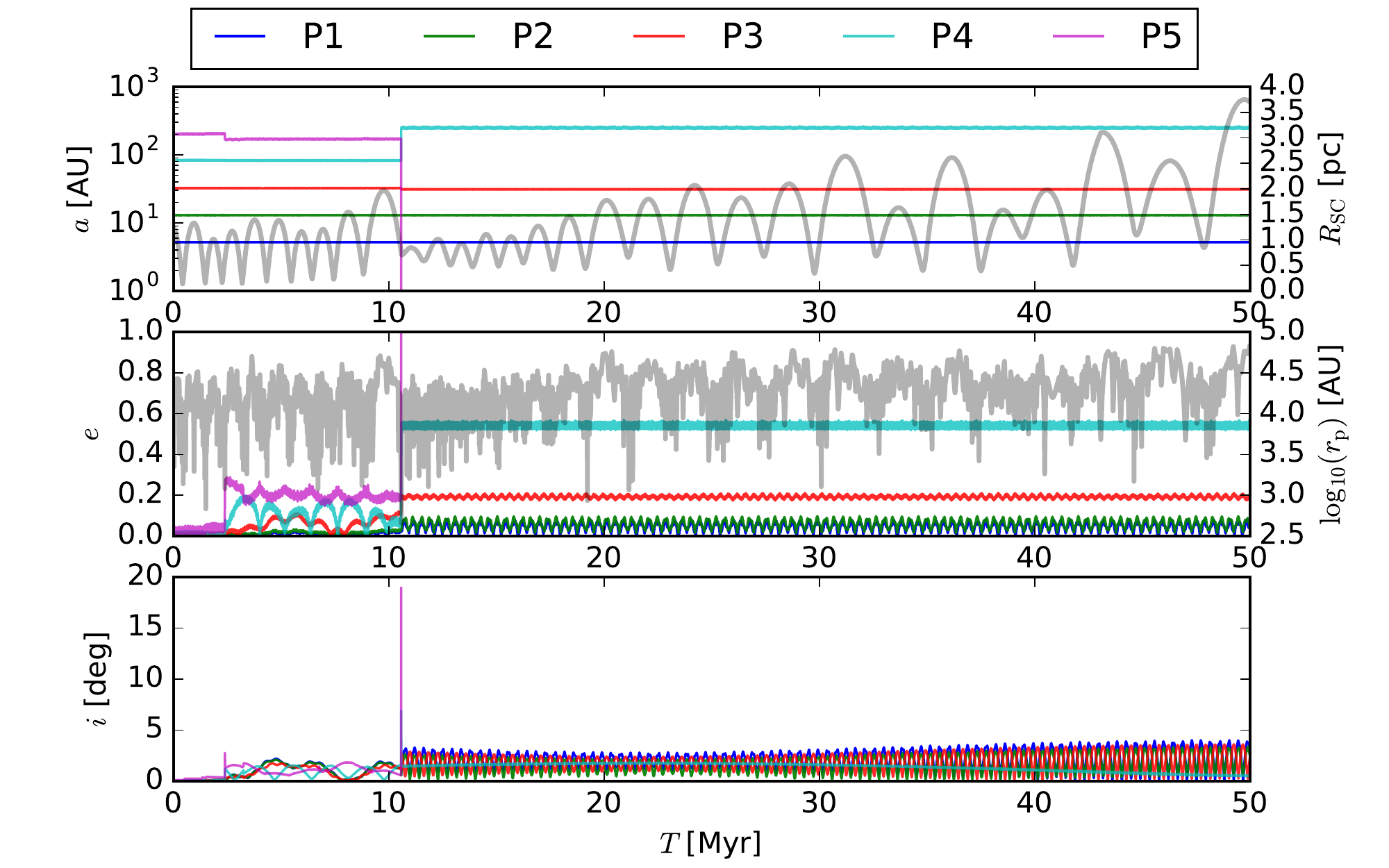}
\caption{Simulation of a planetary system perturbed by stellar encounters in star cluster environment. Planets in this system are equal mass (1 Jupiter mass), arranged in initially circular coplanar orbits. The semi-major axis, eccentricity, and inclination of each planet are plotted as a function of time on the top, middle and bottom panel, respectively. At the top panel, the thick gray curve shows the distance from the planetary system to the cluster center (in parsec); in the middle panel, the thick gray curve shows the distance of the closest perturbing star (in AU, log scale). The planetary system is destabilized by a close encounter at $T\sim 3$~Myr. A subsequent close encounter at $T \sim 11$~Myr causes the ejection of the outermost planet P5 and also excites P4 to $e \sim 0.6$, which in turn results in stronger planet-planet interactions. Because planetary systems are chaotic few-body systems, their stability can only be derived statistically from an ensemble of simulations. Each simulation takes $\sim 30$~hours to finish on a modern CPU core using the \texttt{IAS15} \citep{2015MNRAS.446.1424R} integrator, which is available in the \texttt{rebound} \citep{2012A&A...537A.128R} package. Our parameter space consists of three different models of star clusters and four different architectures of planetary systems, making it 12 ensembles of simulations. Each ensemble contains 100 planetary systems individual simulations, and therefore the total number of simulations is 1200. \simon{} distributes these simulations to 12 computing nodes (16 CPUs per node) and schedules them dynamically for optimal load balancing. }
\label{fig:p_sys_sc}
\end{figure*}

The simulations are performed in several stages. First, the host cluster is simulated without planets, and the simulation data are stored at a time resolution of $10^3$ year \citep{Cai:2015aa}. In the second stage, a fraction of stars in the host cluster are selected as the host for a planetary system. Based on the stored cluster data, the positions and velocities of the stars in the vicinity of the planetary system are calculated; these are considered the perturbers of the targeted planetary system. In the third stage, we evolve each planetary system while with the perturbation due to the pre-calculated encounter history taken into account while integrating the equations of motion. An simulation example is presented in Figure~\ref{fig:p_sys_sc} (full scientific results are presented in \citealt{2017arXiv170603789C}). 
It would have been very hard to perform these simulations without a tool such as \simon{}.

\subsection{Parallel Observational Data Reduction: Planet Detection in an Extrasolar Ring System in the Sco-Cen OB Association}

The workflow was originally designed for farming simulations, but in this example we use it for supporting the observational data reduction of an ongoing project. 

In April 2007 the star  1SWASP J140747.93-394542.6 in the OB association Sco-Cen OB was observed for 56 days to have a complex light curve with a number of unexplained features. Based on the emergent pattern it was deduced that a series of dips could be explained by a massive extrasolar ring system with a radius of 0.6~AU \citep{2015ApJ...800..126K}. To confirm the detection of this planet new observations were scheduled in March 2016 with the Very Large Telescope using the Zurich Imaging POLarimeter (ZIMPOL). The reason ZIMPOL is used is that the massive ring system is expected to act like a large mirror reflecting and polarizing light from the host star in our direction. Given that J1407 produces unpolarized light and the reflected light from J1407b will be partially polarized, Polarimetric Differential Imaging \citep[PDI, see][]{2016arXiv161006609D} should be able to filter out the starlight, leaving a polarized signal produced by the companion J1407b.

Unfortunately, the planet was not found using PDI from the data obtained, but the available data presents an interesting opportunity to investigation the 4-dimensional parameter space. This is done by injecting a modified copy of the stellar point-spread function at different locations. This results in four parameters, two spatial coordinates (angular distance and position angle), the intensity of reflected light (expressed as a magnitude difference) and the degree of linear polarization. The parameter space consists of $\sim 2.4 \times 10^5$ data points, and we adopted \simon{} to automate the parameter space coverage search.

\subsection{Pulsar survey tasks management: an essential chain of FAST telescope data processing pipeline}

\simon{} is also useful in processing huge amounts of data generated from large sky surveys, such as the pulsar survey using the Five-hundred-meter Aperture Spherical Radio Telescope (FAST)\footnote{\href{http://fast.bao.ac.cn/en/}{\tt http://fast.bao.ac.cn/en/}}. FAST is the world largest single dish telescope recently built in China. Construction of FAST was completed in Sept. 2016, and scientific observations will start after a successful commissioning period. One of the key science project being planned is an all-sky pulsar survey using a 19-beam feedhorn array receiver covering the frequency range of 1.05-1.45~GHz \citep{2011IJMPD..20..989N}. Such survey is expected to discover thousands of new pulsars previously unknown. However, searching for pulsars require extremely high computing power and huge amounts of data are expected to be generated from the FAST survey. The sampling rate is 20000 times per second, and the data is recorded in a 16K channel digital backend. Thus each of the beam receivers will generate 80~MB data per second. Currently, the data are recorded to a file every 30 seconds, and the size of each file is 2.4~GB. The 19 beam receiver will generate 2280 files per hour and 22800 files per 10 hours nights. Each file will be processed with the PRESTO suite of software \citep{2002AJ....124.1788R} to search for periodic signals and identify pulsars. A cluster of computers will be used to process the data in a parallel computing environment. So there will be more than several hundred of processes running each time, and each of these processes can generate up to tens of pulsar candidates. Some of these candidates are from radio frequency interference (RFI) which need to be checked by either human or artificial intelligence. Therefore, intensive bookkeeping is required to keep track of all the data files, processes, RFIs and pulsar candidates, together with the related pulsar parameters. Such bookkeeping was handled manually in traditional pulsar searching when the data rate was low, but it becomes impossible in the FAST pulsar data processing. We are developing a data reduction pipeline to automatically process the FAST data and \simon{} will be playing a critical role in tracking all the pulsar search processes and data files as well as recording the pulsar candidates and parameters.

\section{Conclusion}
\label{sec:conclusion}
We present \simon{}, the Simulation Monitor and driver for simulation and data processing in computational astrophysics. 
The package is lightweight, easy to use, publicly available and implemented using the \texttt{Python} programming language.

Its development is a direct response to the challenges of managing large ensembles of prolonged numerical simulations in a high-performance distributed computational environments. The package is designed to automatically handle an arbitrary number of simulations running with a variety of codes. Although our primary development was aimed at the use of gravitational $N$-body codes within the {\tt AMUSE} \citep{2009NewA...14..369P,2013CoPhC.183..456P,2012ASPC..453..129M,2013A&A...557A..84P} framework, it can be used to monitor, distribute and handle any astrophysical code, including simulation codes and observational data pipelines. The environment supports a daemon mode and an interactive mode. When running in the daemon mode, the real-time status of all managed simulations is gathered automatically, scheduled according to the machine capacity, takes care of the bookkeeping, and performs post-processing tasks with minimum human supervision. When running in the interactive mode, it provides a dashboard with an overview of the status for all managed simulations, allowing the user to manually control individual simulations. 

While \simon{} is an automatic workflow for computational astrophysics aiming to minimize human supervision, it is not designed to eliminate the need of human supervision. In particular, numerical difficulties that can be fixed by simply restarting the code and/or changing some timestep parameters are a minority. For this reason, \simon{} allows the users to define a threshold at which maximum attempts of automatic restarts are tolerated. If a simulation is restarted more times than this threshold, more profound issues are likely exist in the numerical code. In this case, \simon{} marks the simulation as ``ERROR'', notifies the users and will no longer handle this particular simulation. 

The development of \simon{} is motivated by our growing demand on automated simulations, but it applies equally well to other fields in which researchers are overwhelmed by the number of simulations or data processing tasks. Although it was originally developed for astrophysical $N$-body simulations, we hope to serve the wider astrophysics community by making this useful tool available.

\section*{Acknowledgements}
We thank the anonymous referee for his/her comments that helped to improve the manuscript. All authors appreciate the invaluable input from Dirk van Dam and Christian Ginski, and their test of \simon{} with their observational data reduction pipeline.
This work was supported by the Netherlands Research Council NWO (grants \#621.016.701 [LGM-II]) by the Netherlands Research School for Astronomy (NOVA). This research was supported by the Interuniversity Attraction Poles Programme (initiated by the Belgian Science Policy Office, IAP P7/08 CHARM), the European Union's Horizon 2020 research and innovation programme under grant agreement No 671564 (COMPAT project), and National Natural Science Foundation of China (NSFC No. U1531246). PXQ acknowledges support by the China Scholarship Council (CSC) and the funds from Professor Alyssa A. Goodman (CfA/Harvard).



\appendix
\section{Supporting Tables}
\label{sec:support_tables}

\begin{table*}[ht]
\centering
\caption{General attributes of a simulation task. The list of attributes may differ from the actual numerical codes, but they can be extended through a Python \texttt{dict} data structure.}
\begin{tabular}{lll}
\hline
\hline
{\bf Category} & {\bf Properties} & {\bf Example}\\
\hline
Paths & Configuration file, input files, output directory  & {\tt /path/to/data/dir} \\
Type  & The numerical code used to carry out the simulation & {\tt NBODY6}, and all codes supported by {\tt AMUSE}\footnote{\href{http://amusecode.org}{\tt http://amusecode.org}}  \\
Model & Start/Termination criteria, current model time     & $t=5, t_{\rm start}=0, t_{\rm end} = 10$  \\
Process & Process ID, process launch timestamp             & {\tt PID=12345}                     \\
Commands & Commands to start/restart/stop a simulation     & {\tt ./simulation\_code}             \\
Relation & IDs of the parental simulations and sub-simulations & {\tt sim\_id=2, parent=1, children=[5,6,7]} \\ 
Status   & Current status of the simulation                & {\tt RUN/STALL/STOP/DONE/ERROR}   \\
\hline
\hline
\end{tabular}
\label{tab:sim_attr}
\end{table*}

\begin{table*}[h]
\caption{Supported manual actions in interactive mode}
\label{tab:interactive_mode_task_list}
\centering
\begin{tabular}{ll}
\hline
\hline
Task name& Description\\
\hline
List Simulations & Generate a status overview of all managed simulations \\
Select Simulations & Allows the users to select multiple simulation and execute command in batch\\
New Run & Start new simulations from beginning point\\
Restart & Restart the simulation from crashing point\\
Check status & Check the recent or current calculation results and print it\\
UNIX shell & Execute an UNIX shell command in the simulation directory\\
Stop Simulations & Send a stop request to the simulation code \\
Delete Simulations & Delete the simulation instance and all its substance\\
Kill Simulations & Kill the UNIX process associate with a simulation task\\
Backup Restart File & Backup the simulation checkpoint files (for restarting purpose in the future)\\
Post Processing & Perform (post)-processing (usually) after the simulation is done\\
Quit & Quit the \simon{} interactive mode \\
\hline
\hline
\end{tabular}
\label{tab:task_list}
\end{table*}

\begin{table*}[h]
\centering
\caption{A list of generic methods used for controlling an arbitrary simulation. \simon{} provides general-purpose implementation of these methods, but the actual behavior of these methods can either be defined in the configure file (using shell commands; Python programming not required) or be overridden by the a code-specific module (Python programming required).}
\begin{tabular}{ll}
\hline
\hline
{\bf Name of the abstract method} & {\bf Description} \\
\hline
\texttt{sim\_init()}          & Perform necessary initialization procedures in order to start the simulation. \\
\texttt{sim\_start()}         & Start the simulation   \\
\texttt{sim\_restart()}       & Restart the simulation \\
\texttt{sim\_get\_status()}   & Get the current status of the simulation \\
\texttt{sim\_stop()}          & Stop the simulation using the mechanism provided by the code \\
\texttt{sim\_kill()}          & Kill the simulation process forcibly (when the code stalls) \\

\texttt{sim\_backup\_checkpoint()}  & Backup the restart files \\
\texttt{sim\_delete()}        & Delete the simulation data \\
\texttt{sim\_clean()}         & Clean up the simulation data, except for the input files and restart files \\
\texttt{sim\_reset()}         & Reset the simulation, leaving only the input files \\
\texttt{sim\_shell\_exec()}   & Execute a UNIX shell command on the simulation data directory \\
\texttt{sim\_finalize()}      & Finalize the simulation (e.g. perform data processing) after the simulation is finished. \\
\hline
\hline
\end{tabular}
\label{tab:sim_func}
\end{table*}



\end{document}